# Optimisation of Power Modulation for Hall-Héroult Cells: Process Operability and Constraints as Virtual Energy Storage

Choon-Jie Wong[a], Adam A. Larkin[a], Jie Bao[a,†], Maria Skyllas-Kazacos[a], Barry J. Welch[a], Nadia Ahli[b], Maitha Faraj[b], Mohamed Mahmoud[b]

[a] *School of Chemical Engineering, University of New South Wales, Sydney, Australia.*

[b] *Emirates Global Aluminium, Jebel Ali Operations, P.O. Box 3627, Dubai, United Arab Emirates.*

† *Corresponding Author (email: j.bao@unsw.edu.au).*

**Abstract**

Aluminium is manufactured through the Hall-Héroult process, which is very energy intensive. Power modulation, as an industrial-scale demand-side power management approach, allows aluminium smelters to operate with variable power consumption rates and as such be powered by renewable energy sources. In this way, aluminium smelting cells can be used as a large virtual energy storage to balance power demand-supply and stabilise electrical grids. This paper studies the potential optimal power modulation operating conditions, including time-varying line current and anode-cathode distance (ACD) profiles to maximise the aluminium reduction cell profitability subject to constraints on the cell thermal balance. To deal with the complex cell dynamics which are spatially distributed and multi-timescale, a novel optimisation approach that utilises both reduced-order and detailed models is developed. The results yield insight into the optimal line current and ACD profiles for different power modulation scenarios including the time of use electricity tariff and spot price. These results can form the foundation for further studies into online control policies of aluminium reduction cells.

*Keywords:* Virtual energy storage, Demand-side power management, Hall-Héroult process, Optimisation, Power modulation.

## 1. Introduction

The world is currently pushing for a greener, low-carbon, and more sustainable future. The introduction of alternative/renewable energy sources, in general, has posed challenges for electrical grid operators who must also ensure network stability. With greater penetration of variable renewable energy, such as solar or wind energy, the instantaneous generation capacity has become subjected to seasonal and diurnal variations. In addition, the shift from traditionally stable fossil fuels baseload in large interconnected urban areas has widen the gap between power generation and instantaneous demand, leading to increased frequency of large-scale brownouts or blackouts and high variability in spot market electrical prices [1-5]. It is imperative that electrical grids remain stable and reliable during the transition toward greener and more sustainable sources of energy to meet net emission targets and maintain stable electricity prices.

Aluminium is an important material that plays an important role in these decarbonisation efforts. Due to its properties such as high strength-to-weight ratio, infinite recyclability, and its formability into complex modern shapes, aluminium is currently among metals that have the fastest growing demands [6]. This is mainly driven by the transport sector that requires lighter cars, trains, and planes to help improve fuel efficiency. As such, aluminium production (primarily with the Hall-Héroult process) has been increasing over the years. However, this process is highly energy intensive, constituting a global energy consumption of nearly one million gigawatt-hours annually [7]. In addition, the aluminium smelting industry constitutes a large proportion of national power consumption. For example, the Australian aluminium industry consumes around 14% of its energy generation capacity [8]. Therefore, there is a strong incentive to produce low-carbon "green" aluminium with renewable energy sources.

An opportunity to address both challenges described above is provided by the massive power requirements of the Hall-Héroult process, coupled with the large thermal capacity of each aluminium reduction cell. These cells can function as "virtual energy storage" by modulating their power consumption [9, 10]. By enabling flexible power consumption of aluminium smelters, these large energy consumers can also provide significant demand-side response services to electrical grids to balance the demand-supply mismatch of the grid [11, 12]. At the same time, this allows for a direct response to the energy market, hence helps the smelters to avoid the high electricity cost associated with times of peak demand on the grid and potentially provides them an additional income stream through demand-side response services and making an arbitrage profit [5]. Power modulation of aluminium smelter cells can be achieved by adjusting either the line current or voltage at which the cells are operated. This enables industrial scale demand-side power management approach that provides flexibility in using and accommodating the variations in renewable energy sources. This allows manufacturing of low-carbon aluminium with cleaner energy sources.

Virtual energy storage achieved via demand-side management is a widely recognised strategy in other energy-intense industries, including the production of steel [13-16], chlorine [17-19], and cement [20, 21], as well as in seawater desalination [22-24], pulp and paper industry [25, 26], air separation plant [27], and job scheduling in data centres [28]. However, the power modulations of aluminium reduction cells are not trivial, as the delicate mass and thermal balances inside the cells are dependent on the current, voltage, and materials supplied to the cell. The variation of cell voltage primarily alters the rate of heat generation, while the current also drives the cell electrochemistry and determines the feed consumption rate and metal production rate. Thus, this paper develops a model-based economic optimiser that determines the optimal current and cell voltage (achieved by the manipulation of anode-cathode distance, or ACD) trajectories which maximises smelter operation profitability while maintaining cell mass and thermal balance for operational feasibility.

The process variables in the aluminium reduction process exhibit multi-timescale spatial and temporal variations [29], which significantly complicates modelling and increases computational complexity. Additionally, the optimiser requires a large time horizon that aligns with the cycles of renewable energy production or energy cost. This work proposes a novel approach to manage the computational complexity by employing a feedback-update philosophy. In this approach, a spatially distributed cell model is combined with a reduced-order model in a feedback loop to compute optimal regimes efficiently, using a shrinking time horizon for optimisation. This method can be integrated with current advances in power modulation, such as the shell-heat-exchanger, to further enhance economic benefits for the aluminium smelting industry and electrical grids through demand-side management.

## 2. Power Modulation of Aluminium Smelter Cells

Figure 1 shows the structural design of a typical modern aluminium reduction cell. Modern cells typically use 16–56 prebaked carbon anodes suspended in an electrolyte (molten cryolite with other additives) at temperatures typically between 955–970 °C [30, 31]. The highly corrosive nature of molten electrolyte requires a layer of frozen cryolite (known as ledge) at the boundary of the cell cavity, preventing the electrolyte from damaging the sidewall interior and shortening the cell lifespan [32]. Hence, any power modulation performed on the cells must carefully balance the aluminium metal production for economic benefit and cell thermal balance such that there is sufficient molten electrolyte without melting away the protective cell frozen ledge [32].

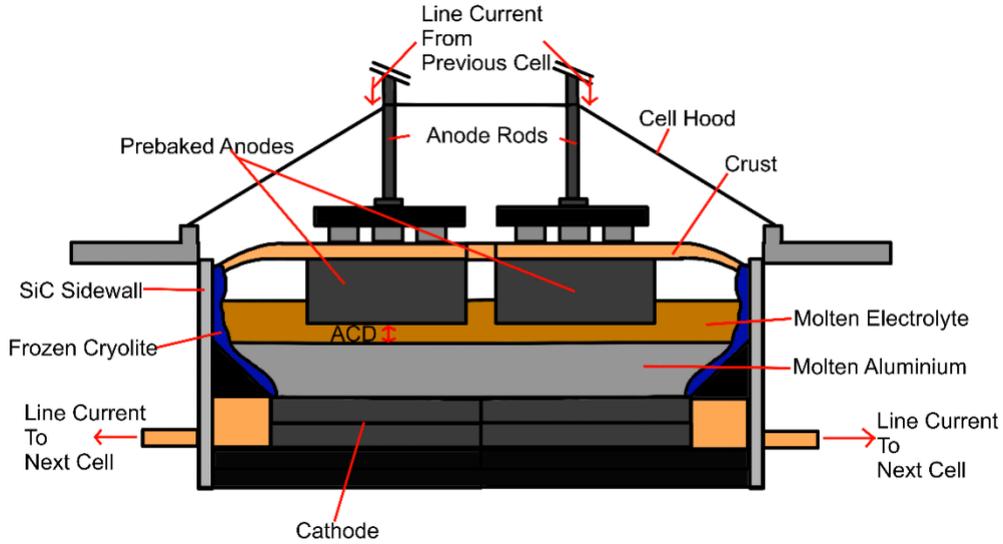

Figure 1: Cross-sectional representation of a prebake anode aluminium reduction cell.

Power modulation is achieved by adjusting the regulated line current (*i.e.,* the electrical current sent through multiple cells connected in series) and the ACD, which influences the ohmic heat generation through varying electrical resistance. As the ACD determines the volume of electrolyte through which the current flows, it directly affects the cell's electrical resistance and, consequently, the cell voltage. Figure 2 shows extreme operational strategies for power modulation. In the first strategy (Figure 2a), both line current and ACD are varied concurrently to maintain cell thermal balance, ensuring that the heat generation component remains constant despite fluctuations in power input; Figure 3 shows a calculation of how ACD may be manipulated at different line currents to achieve constant heat generation. In contrast, the second strategy (Figure 2b) involves modulating only the line current, thereby exploiting the cell's thermal capacity to absorb thermal imbalances. This latter approach allows for maximum flexibility in power modulation but increases reliance on the cell's ability to tolerate transient thermal deviations. In practice, viable operating strategies lie between these two extremes, where the operability of power deviation depends on both its magnitude and duration, as well as the cell design and condition.

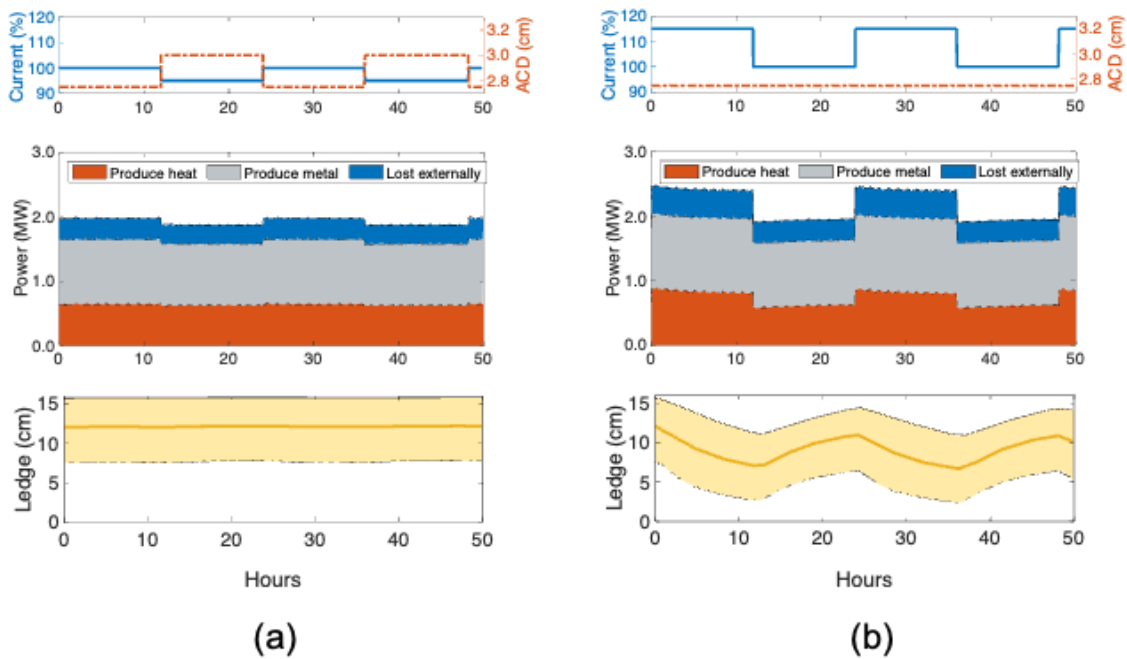

Figure 2: Two limiting operational strategies for power modulation. (a) Full compensation of cell thermal balance, (b) Full utilisation of cell thermal capacity.

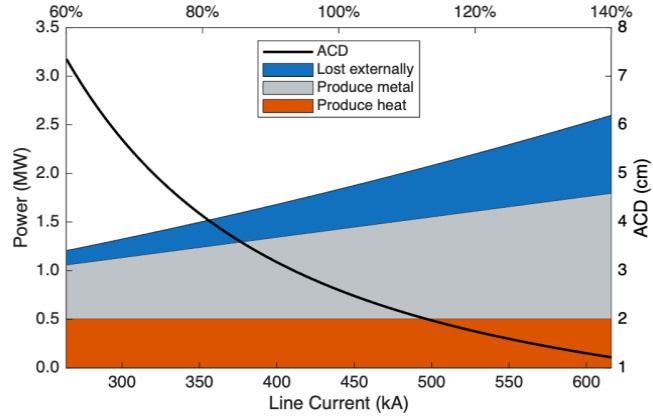

Figure 3: ACD adjustment required to ensure constant heat generation rate at different line currents.

While the trend to build larger cells with increasing line current (exceeding 400 kA [9, 33, 34]) and decreasing feeder-number-to-current ratio has worsened the spatial non-homogeneity in the cell electrolyte [35], the size increase however imparts substantially large thermal capacities. This presents an opportunity for the industrial process to act as "virtual energy storage" and participate in demand-side-response, *i.e.*, the process can accommodate significant dynamical changes in power consumption without threatening the cell thermal balance under appropriately determined line current and ACD trajectories.

Certain forms of power modulation have been trialled on industrial aluminium reduction cells, as early as in the 1970s following the oil crisis. Valesul Alumínio (1985) and Alcanbrasil (1987) reported implementations of power modulation, while other correspondences revealed that poor thermal balance control had led to cryolite freezing in the cell bottom, which disturbed the electrolyte and metal flows and led to instability of cell operation and losses in process efficiency [36]. Similarly, power price increases in Germany since the energy market's liberalisation and replacement of nuclear generation with renewable power has motivated the implementation of power modulation strategies in German aluminium smelters, such as TRIMET Aluminium [9, 37, 38]. TRIMET used simple power modulation regimes involving set percentage increases or decreases in line current inputs to the cell, with some experiments being subject to a compensatory cell voltage, through ACD adjustment, to meet thermal balance requirements [9]. Shell-heat-exchanger technologies [2, 39, 40] were also introduced in other smelters primarily to accommodate greater magnitudes of fixed power increases/decreases (load shifting). However, dynamically optimised trajectories of line current and ACD was not employed to maximise potential economic benefit during operation and consequences to energy efficiency through wasted additional heat drawn from the cells [33].

Whilst the literature presents pragmatic solutions leading to feasible operation, there are opportunities for attaining tighter cell control while maximising economic benefit during power modulation via dynamically optimising the operation regimes. This involves developing a model-based optimiser to determine the economically optimal power modulation regimes, subject to thermal balance constraints (such as maintaining ledge thickness) to ensure operational feasibility. This work includes the use of feedback-update philosophy to address the computational complexity of optimisation, since very large models are needed to deal with the spatial distribution of thermal and material properties of aluminium reduction cells, as well as to deal with the multiple timescales of the process.

## 3. Model-Based Optimisation Framework

This paper is scoped to the analysis of economic feasibility and process operability under power modulation conditions. Hence, dynamic models were developed to predict the thermal behaviours of an industrial cell, which is representative of all cells in the potline with the same design and technology. Based on these thermal constraints associated with the cell technology, an optimiser is developed to calculate the optimum line current and ACD trajectories (power consumption profile) that maximises economic benefits. However, it is noted that while line current affects all the cells in the potline, the impact of ACD changes is limited to each cell. Therefore, for extensions to optimal real-time control implementation, ACD control should be delegated to a lower-level controller that considers the stability and conditions of each cell; this is outside the scope of the current feasibility study.

As the dynamics of the smelting process are spatially distributed, a spatially distributed dynamic model is required to accurately compute the spatial and temporal process variables. In terms of control degree of freedom, the manipulated variables are line current and ACD, which effect change across the entire cell since they cannot be controlled for each anode. Therefore, a reduced-order model with appropriate complexity was developed to compute temporal process dynamics, so large optimisation windows of several days in length can be computed efficiently. To maintain overall optimisation accuracy, we introduce a novel feedback-update technique where accurate states from the spatially distributed model are iteratively used to update those in the reduced-order model as the optimisation proceeds.

### 3.1 Dynamic Cell Models

#### 3.1.1 Spatially Distributed Mass and Thermal Model

A distributed dynamic model of modern aluminium smelter cells with coupled material and thermal balance derived from the first principles was published in our previous works [41, 42]. It can compute the local cell dynamics, such as those of the electrolyte composition and temperature, based on consumption of materials and generation of heat at various spatial locations. As a dynamic model based on integrated material and energy balances, it is suitable for various operating conditions including power modulations, as it also calculates the cell voltage and distribution of anode currents based on the changing line current and local cell conditions – the latter of which governs the local rate of the electrowinning process (including the rates of heat generation and material consumption).

The spatial discretisation in this spatially distributed dynamic model is conducted according to cell design and the anode geometry and locations. For a 36-anodes industrial cell studied in this work (see Figure 4), the cell cavity, ledge, silicon carbide sidewall, and the steel shell are discretised into an 11-by-43 grid, where the subsystems may not share the same grid dimensions, *i.e.*, the volume and mass in each grid varies. This spatially distributed model captures the spatial and temporal dynamics of the tightly coupled material and thermal properties. It accounts for local electrolyte mass concentrations, local temperatures, local ledge thickness and electrolyte height – amounting to greater than 800 state variables.

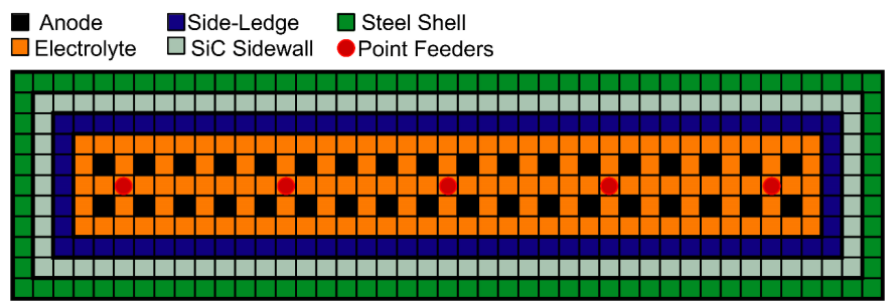

Figure 4: Thermal components in the spatially distributed dynamic model.

### 3.1.2 Reduced-Order Thermal Model

The optimiser requires a dynamic cell model with appropriate computational complexity to optimise their power consumption subject to the thermal constraints of the cell design and technology. As line current and ACD variations affect the entire cell, reduced-order lumped parameter thermal model can be used to predict the temporal thermal behaviour of cells in response to power variations. The dynamics of individual alumina dumps can also be simplified by assuming perfect alumina feeding at the base feed rate which subsequently dissolves completely. This is because on the longer timescale that thermal balance dynamics exist, the dynamics of mass balance disappears.

*Electrolyte temperature.* The thermal balance for the electrolyte (bath) can be represented as follows:

$$c_{\text{bath}} \frac{d(m_{\text{bath}} T_{\text{bath}})}{dt} = Q_{\text{gen}} - h_{\text{ldg-bath}} A_{\text{ldg-bath}} (T_{\text{bath}} - T_{\text{liq}}), \tag{1}$$

where $T_{\text{bath}}$ is the temperature of the electrolyte with mass $m_{\text{bath}}$ and specific heat capacity $c_{\text{bath}}$. $h_{\text{ldg-bath}}$ is the convective transfer coefficient between the electrolyte and frozen ledge, with interfacial area $A_{\text{ldg-bath}}$. $Q_{\text{gen}}$ is the heat generation rate within the cell, and $T_{\text{liq}}$ is the liquidus temperature, representing the temperature above which the electrolyte is completely liquid.

Ohmic heat generation $Q_{\text{gen}}$ is the product between line current $I_{\text{Line}}$ and cell voltage $V_{\text{cell}}$. However, voltage drop external to the cell $V_{\text{ext}}$ and that due to reagent dissolution, reactions and necessary preheating (6.6 kWh/kg Al [35]) do not contribute towards heat generation and should be subtracted:

$$Q_{\text{gen}} = (V_{\text{cell}} - V_{\text{ext}}) \times I_{\text{Line}} - 6600 \dot{m}_{\text{Al}}(I_{\text{Line}}), \tag{2}$$

where $\dot{m}_{\text{Al}}$ is the production rate of aluminium metal given by Faraday's law of electrolysis:

$$\dot{m}_{\text{Al}}(I_{\text{Line}}) = \frac{\eta I_{\text{Line}} M_{\text{Al}}}{zF} \times 3600, \tag{3}$$

with $\eta$ representing the current efficiency, $M_{\text{Al}}$ is the molecular weight of aluminium, $z$ is the number of electrons transferred and $F$ is the Faraday constant.

*Ledge thickness.* With the protective layer of frozen cryolite ledge at the boundary of the cell cavity, ledge is assumed to dissolve or form at the liquidus temperature. The formation or dissolution is driven by the difference of heat transferred from the bath to the ledge surface (at liquidus temperature) $Q_{\text{bath-ldgb}}$, and that from the ledge surface to the ledge $Q_{\text{ldgb-ldg}}$:

$$\frac{d(m_{\text{ldg}})}{dt} = \frac{Q_{\text{ldgb-ldg}} - Q_{\text{bath-ldgb}}}{\Delta H_f} \tag{4}$$

$$Q_{\text{ldgb-ldg}} = \frac{k_{\text{ldg}}(A_{\text{ldg-bath}} + A_{\text{ldg-met}})}{0.5 l_{\text{ldg}}} (T_{\text{liq}} - T_{\text{ldg}}) \tag{5}$$

$$Q_{\text{bath-ldgb}} = [h_{\text{ldg-bath}} A_{\text{ldg-bath}} + h_{\text{ldg-met}} A_{\text{ldg-met}}](T_{\text{bath}} - T_{\text{liq}}), \tag{6}$$

where $m_{\text{ldg}}$ is the mass of the ledge, with temperature $T_{\text{ldg}}$ and thermal conductivity coefficient $k_{\text{ldg}}$. $h_{\text{ldg-met}}$ is the heat convection coefficient between ledge and aluminium, with interfacial area $A_{\text{ldg-met}}$. $\Delta H_f$ is the enthalpy of fusion for the ledge material. Thus, the ledge thickness $l_{\text{ldg}}$ (assumed to be uniform in this lumped parameter model) can be calculated:

$$l_{\text{ldg}} = \frac{m_{\text{ldg}}}{\rho_{\text{ldg}}(A_{\text{ldg-bath}} + A_{\text{ldg-met}})}, \tag{7}$$

where $\rho_{\text{ldg}}$ is density of the ledge. The interfacial areas also vary according to ledge thickness, *e.g.*:

$$A_{\text{ldg-bath}} = 2\left((L - 2l_{\text{ldg}}) + (W - 2l_{\text{ldg}})\right)H, \tag{8}$$

where $L$ and $W$ are the length and width of the cell cavity respectively, and $H$ is the ledge height.

*Electrolyte mass.* Given that formation and dissolution of ledge transfers mass to or from the electrolyte, the change in electrolyte mass is:

$$\frac{d(m_{\text{bath}})}{dt} = -\frac{d(m_{\text{ldg}})}{dt} = -\rho_{\text{ldg}}\left(A_{\text{ldg-bath}} + A_{\text{ldg-met}}\right)\frac{d(l_{\text{ldg}})}{dt}. \tag{9}$$

*Ledge temperature.* The ledge temperature dynamics is driven by the difference in heat transfer from ledge surface (at liquidus temperature to the ledge, and from the ledge to the sidewall materials:

$$c_{\text{ldg}}\frac{d(m_{\text{ldg}}T_{\text{ldg}})}{dt} = Q_{\text{ldgb-ldg}} - \frac{(T_{\text{ldg}} - T_{\text{sw}})}{R_{\text{ldg-sw}}}, \tag{10}$$

where $c_{\text{ldg}}$ is the specific heat capacity of the ledge. The sidewall has temperature $T_{\text{sw}}$, thermal conductivity $k_{\text{sw}}$ and thickness $l_{\text{sw}}$. $R_{\text{ldg-sw}}$ is the thermal resistance from frozen ledge to sidewall:

$$R_{\text{ldg-sw}} = \frac{1}{A_{\text{ldg-sw}}}\left(\frac{0.5\, l_{\text{ldg}}}{k_{\text{ldg}}} + \frac{0.5\, l_{\text{sw}}}{k_{\text{sw}}}\right). \tag{11}$$

The liquidus temperature depends on the chemical composition of the cell electrolyte. Its value in degree Celsius is modelled based upon the empirical relation to wt/wt concentrations of alumina ($Al_2O_3$), aluminium fluoride ($AlF_3$), calcium fluoride ($CaF_2$), lithium fluoride ($LiF$), magnesium fluoride ($MgF_2$) and potassium fluoride ($KF$) [43]:

$$\begin{aligned}T_{\text{liq}} =\\ 1011 + 0.5\,\%AlF_3 - 0.13\,(\%AlF_3)^{2.2} - \frac{3.45\,\%CaF_2}{1 + 0.0173\,\%CaF_2}\\ + 0.124(\%CaF_2)(\%AlF_3) - 0.00542\left((\%CaF_2)(\%AlF_3)\right)^{1.5}\\ - \frac{7.93(\%Al_2O_3)}{1 + 0.0936(\%Al_2O_3) - 0.0017(\%Al_2O_3)^2 - 0.0023(\%AlF_3)(\%Al_2O_3)}\\ - \frac{8.90(\%LiF)}{1 + 0.0047(\%LiF) + 0.0010(\%AlF_3)^2} - 3.95(\%MgF_2) - 3.95(\%KF).\end{aligned} \tag{12}$$

*Sidewall temperature.* The dynamics in the temperature of the sidewall can be represented as follows:

$$m_{\text{sw}}c_{\text{sw}}\frac{d(T_{\text{sw}})}{dt} = \frac{(T_{\text{ldg}} - T_{\text{sw}})}{R_{\text{ldg-sw}}} - \frac{(T_{\text{sw}} - T_{\text{amb}})}{R_{\text{sw-amb}}}, \tag{13}$$

where $T_{\text{sw}}$ is the temperature of the sidewall, with mass $m_{\text{sw}}$ and specific heat capacity $c_{\text{sw}}$. $T_{amb}$ is the ambient surroundings temperature. $R_{\text{sw-amb}}$ represents the lumped-parameter thermal resistance from the centre of the sidewall layer to the surroundings:

$$R_{\text{sw-amb}} = \frac{1}{A_{\text{ext}}}\left(\frac{0.5 l_{\text{sw}}}{k_{\text{sw}}} + \frac{l_{\text{shell}}}{k_{\text{shell}}} + \frac{1}{h_{\text{shell-amb}}}\right), \tag{14}$$

where $A_{\text{ext}}$ is the external area of the cell in contact with the surroundings, $k_{\text{shell}}$ is the conductivity of the surrounding steel shell, and $h_{\text{shell-amb}}$ is the convective coefficient of transfer between the cell exterior and ambient air.

*Cell voltage.* The cell voltage equation has been published in the literature [41, 42, 44]. The following voltage components are included in the model:

$$V_{\text{cell}} = V_{\text{rev}} + \eta_{\text{sa}} + \eta_{\text{ca}} + \eta_{\text{cc}} + I_{\text{Line}}(R_{\text{bath}} + R_{\text{bub}} + R_{\text{an}} + R_{\text{ca}} + R_{\text{ext}}), \tag{15}$$

where $V_{rev}$ is the reversible potential, $\eta_{sa}$ is the surface overvoltage at the anode, $\eta_{ca}$ is the concentration overvoltage at the anode, $\eta_{cc}$ is the concentration overvoltage at the cathode. $IR_{bath}$, $IR_{bub}$, $IR_{an}$, $IR_{ca}$, and $IR_{ext}$ are the voltage drops across the electrolyte, bubble layer, anode, cathode, and external cell superstructures respectively.

The electrolyte resistance $R_{bath}$ is dependent on the electrolyte composition as well as the thickness of electrolyte layer (*i.e.,* ACD):

$$R_{bath} = \frac{(D - d_b)}{\kappa_{bath} A_{an}}, \tag{16}$$

where $\kappa_{bath}$ is the electrolyte conductivity. $D$ is the anode cathode distance; it is emphasised that ACD is a manipulated variable to be optimised, and its direct impact on cell voltage is observed here. All these terms are also directly or indirectly impacted by line current.

*Model state equations and output equation.* Summarising the equations above, the state equations for spatially averaged electrolyte temperature, ledge temperature, sidewall temperature, ledge mass, and ledge thickness are given by:

$$c_{bath} \frac{d(m_{bath} T_{bath})}{dt} = Q_{gen} - h_{ldg-bath} A_{ldg-bath} (T_{bath} - T_{liq}) \tag{17}$$

$$c_{ldg} \frac{d(m_{ldg} T_{ldg})}{dt} = \frac{k_{ldg} A_{ldg-sw} (T_{liq} - T_{ldg})}{0.5 \, l_{ldg}} - \frac{(T_{ldg} - T_{sw})}{R_{ldg-sw}} \tag{18}$$

$$m_{sw} c_{sw} \frac{d(T_{sw})}{dt} = \frac{(T_{ldg} - T_{sw})}{R_{ldg-sw}} - \frac{(T_{sw} - T_{amb})}{R_{sw-amb}} \tag{19}$$

$$\frac{d(m_{ldg})}{dt} = \frac{-[h_{ldg-bath} A_{ldg-bath} + h_{ldg-met} A_{ldg-met}](T_{bath} - T_{liq})}{\Delta H_f} \tag{20}$$

$$\rho_{ldg} A_{ldg} \frac{d(l_{ldg})}{dt} = \frac{d(m_{ldg})}{dt} = -\frac{d(m_{bath})}{dt}. \tag{21}$$

The model output is the power usage of a single, representative smelter cell in the potline:

$$P = I_{Line} V_{cell}. \tag{22}$$

### 3.1.3 Discussions and Comparison of Models

To illustrate the similarity and differences between both the spatially distributed and the reduced-order cell models, they are initialised with the same initial conditions and subjected to the same control actions and power modulation condition: a 10% increase in line current from hours 8 to 12. The dynamics in key process variables from both models are compared in Figure 5.

The spatially distributed model was simulated with conventional alumina feeding control and cell resistance control. This model calculates the state trajectories of each spatial location, but only the bounds and weighted mean are plotted for clearer visualisation. This simulation also illustrates the multi-timescale nature of the process; for example, Figure 5(a) shows that the bath temperature takes more than 6 hours to reach steady values, while Figure 5(b) and (c) show that the ledge dynamics take longer than 12 hours. This also demonstrates the large thermal capacity and the long-lasting impact of power modulation, with the transient-state period being similar in length to power modulation periods. This emphasises the need to optimise line current and ACD trajectories throughout the entire period, so the cell thermal constraints are always maintained.

The reduced-order model, despite significant simplifications, predicts similar trends as the spatially distributed model and the magnitudes of key process variables remain well within the bounds. A small difference in response speed is noted between the two models, although its impact is minimal. This arises as the spatially distributed model has spatially discretised subsystems of varying mass, and so its thermal inertia varies spatially. In comparison, the reduced-order model has only one lumped mass parameter for each material, and so its thermal inertia represents the average. The oscillations for electrolyte and ledge temperatures are also absent from the reduced-order model output. This is because the alumina 'demand-feed' pattern has little impact on the cell for the purpose of power modulation which occurs on a much longer timescale. The alumina feeding dynamic was thus left out of the reduced-order model to reduce state dimension and computational complexity.

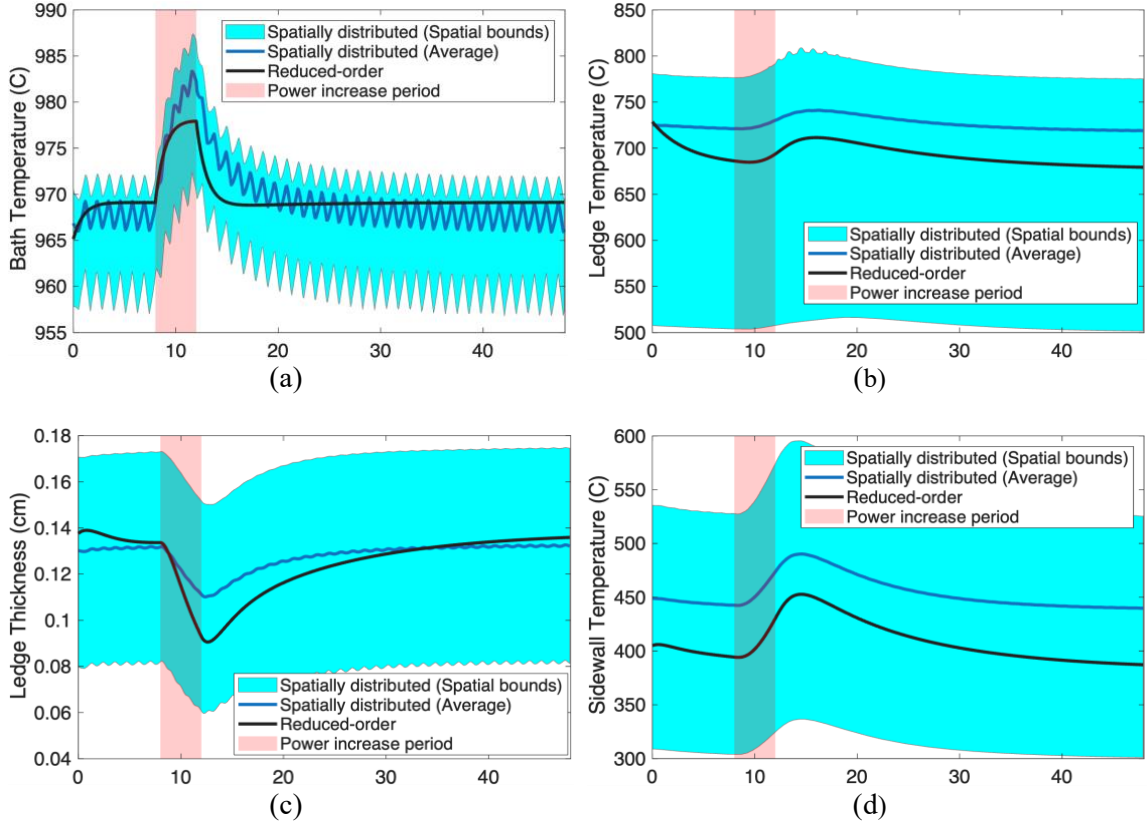

Figure 5: Simulations comparing spatially distributed and reduced-order models subject to a 4-hour 10% increase in line current. Results for (a) electrolyte/bath temperature, (b) ledge temperature, (c) ledge thickness, (d) sidewall temperature.

### 3.2 Nonlinear Program Optimisation with Feedback-Update

The scope of this work is to analyse the economic feasibility and process operability under power modulation. This is conducted by optimising the trajectories of line current and ACD that maximise the economic benefits, while ensuring the cell remains within acceptable operating window (*e.g.*, maintaining the ledge thickness within acceptable bounds). The multi-timescale and spatially distributed nature of material and energy in a cell must be considered. To address these, two solutions are proposed to be used together. The first solution involves formulating the optimisation problem with continuous-time reduced-order model to capture both fast and slow dynamics, avoiding numerical errors and difficulties that arise in discrete-time models due to the multi-timescale dynamics. This problem is then transcribed into a nonlinear program with orthogonal collation, so the constraints can be imposed on the intermediate time points (collocation points) with little discretisation error. The second solution involves a feedback-update strategy, where the reduced-order model is used in optimisation to keep computational complexity low, while periodically updated with feedback from the outputs of the spatially distributed model. This keeps the optimisation model aligned with the full physics and reduces the errors in the reduced-order representation.

### 3.2.1 Optimisation Problem Formulation

The aluminium reduction cell possesses multi-timescale dynamics. For example, Figure 5 shows changes in power input cause mass variables like alumina concentration to respond in minutes, while thermal variables respond slowly over 12 hours. This multi-timescale nature of the process dynamics makes discrete-time models unsuitable for use in dynamic optimisation: fast dynamics require small sampling time periods, which increases the number of decision variables in a time horizon and thus the numerical complexity; slow dynamics require large sampling time, which can cause fast dynamics to be misrepresented (*i.e.*, aliasing). A discrete-time representation may be ill-conditioned which leads to numerical instability. Any discretisation error introduced may also accumulate to unacceptable levels over the long modulation period. Hence, we use continuous-time system model to formulate the optimisation problem to determine the optimal operating condition as a vector function of time, $u(t)$.

The reduced-order model (Eq. 17–21) can be expressed succinctly as:

$$\frac{dx}{dt} = f(x, u), \tag{23}$$

where $x(t)$ is a vector of the state variables: spatial-mean temperatures of electrolyte ($T_{\text{elec}}$), ledge ($T_{\text{ldg}}$), and sidewall ($T_{\text{sw}}$) and spatial-mean of ledge thickness ($L_{\text{ldg}}$), while $u(t)$ is a vector of the input variables: line current and spatial-mean of ACD.

There are several important objectives and requirements to which the optimised power modulation profile must adhere. They can be formulated as:

$$\min_{u(t),\ t_0 \leq t_f} J(t) = \int_{t_0}^{t_f} \left( \text{Expenses}_{\text{cell}}(x, u) - \text{Revenue}_{\text{cell}}(x, u) \right) dt \tag{24a}$$

$$\text{s.t.}$$

$$\text{Eq. 23} \tag{24b}$$

$$x_{\text{lb}} \leq x(t) \leq x_{\text{ub}} \tag{24c}$$

$$u_{\text{lb}} \leq u(t) \leq u_{\text{ub}} \tag{24d}$$

$$\left(\frac{du}{dt}\right)_{\text{lb}} \leq \frac{du(t)}{dt} \leq \left(\frac{du}{dt}\right)_{\text{ub}} \tag{24e}$$

$$-\epsilon \leq x(t_f) - x(0) \leq \epsilon \tag{24f}$$

$$-\alpha \leq P(t) - P_{\text{target}}(t) \leq \alpha. \tag{24g}$$

- Eq. 24a is the economic cost function to be minimised. It is the net revenue per cell from time $t_0$ to $t_f$ ($t_f = 48$-th hour, after two daily modulation cycles), to capture the long-term cell behaviours during power modulation. The expenses are related to costs of raw materials and electricity; the revenue is the market value of aluminium metal produced. It is noted that fixed capital, maintenance, and labour costs are not included in this cost function. Due to other constraints implemented, the impact on pot life [45] is assumed to be negligible.

- Eq. 24b is the smelter cell thermal balance that must be obeyed. It is a continuous-time system model which reduces numerical errors associated with sampling of a multi-timescale system.

- Eq. 24c is the acceptable bounds within which the process variables must remain. For the industrial cell in this study, the average ledge thickness must be between 2 cm (to ensure adequate sidewall protection) and 15 cm (the spacing available between anode and sidewall).

- Eq. 24d is the acceptable bounds within which line current and ACD must remain. The line current (nominal 425 kA) must remain above 200 kA (to meet production requirement), and the ACD (nominal 2.8 cm) must remain between 2.5 and 5 cm (to limit the risk of process faults such as shorting and anodes lifted from bath).

- Eq. 24e is the acceptable bounds for the rate of change of control actions. The line current may not change quicker than 360 kA/h (for process stability), and the ACD may not be adjusted quicker than 0.36 cm/h (to maintain crust integrity).

- Eq. 24f enforces that the ledge thickness at $t = t_f$ (end of two modulation cycles) return to its initial value, to ensure that the cell possesses the thermal capacity for the next power modulation cycles. This requirement is not expressed as a soft constraint (terminal cost) to maintain the physical interpretation of the cost function $J$, which is an economic cost function.

- Eq. 24g, when included in the optimisation problem, enforces that cell power usage $P(t)$ (Eq. 22) matches a target power profile $P_{\text{target}}(t)$, while allowing a small deviation of α. This is useful for Scenario I of optimisation studies, where we vary the power by ±20%.

The above formulation is set up so the trajectories of line current and ACD can be optimised to maximise economic benefits over long-term modulation cycles, while ensuring the cell remains stable. This optimisation problem is transcribed into a nonlinear program in the next section, so it can be solved numerically more efficiently.

### 3.2.2 Nonlinear Programming Problem Formulation

The optimisation problem in the previous section is to be solved with long optimisation period of days. Combined with nonlinear cell dynamics which imposes constraint on the solution, there can be a substantial increase in computational complexity. Inspired by pseudo-spectral optimal control methods [46, 47], we transcribe the optimisation problem into nonlinear program using Lobatto collocation, which approximates the system differential equation as algebraic constraints and integral cost function as weighted sum. The state (*e.g.*, electrolyte temperature) trajectories are approximated as piecewise polynomial functions, and the input (line current and ACD) trajectories as piecewise linear functions. The constraint of system dynamics is exactly imposed on sets of interior collocation points as algebraic constraints and thus, unlike a discrete-time model, does not lead to accumulation of error.

First, partition the time horizon $[t_0, t_f]$ into $M$ segments, as

$$t_0 = 0 \leq t_1 \leq t_2 \leq \cdots \leq t_k \leq t_{k+1} \leq \cdots \leq t_M = t_f. \tag{25}$$

For each time segment $[t_k, t_{k+1}]$, each of length $h$, map the global time $t$ to local time $\tau \in [0, 1]$:

$$t \equiv t_k + (t_{k+1} - t_k)\tau = t_k + h\tau. \tag{26}$$

Choose $N$ Lobatto collocation nodes, at $\{\tau_z\}_{z=0}^{N-1}$. Each state trajectory for the segment $x_k(\tau_z)$ can then be approximated as

$$x_k(\tau_z) = a_{k,0} + a_{k,1}\tau_z + a_{k,2}\tau_z^2 + \cdots + a_{k,N-1}\tau_z^{N-1}, \tag{27}$$

where $a_{k,0}, a_{k,1}, \ldots, a_{k,N-1}$ are the coefficients of the polynomial approximation in the $[t_k, t_{k+1}]$ interval. Noting that $x_k(\tau_z = 0) = a_{k,0}$, the discrete values at the collocation nodes of each state in the $[t_k, t_{k+1}]$ are thus:

$$\begin{bmatrix} x_k(\tau_1) \\ x_k(\tau_2) \\ \vdots \\ x_k(\tau_z) \\ \vdots \\ x_k(\tau_{n-1}) \end{bmatrix} = \begin{bmatrix} x_k(0) \\ x_k(0) \\ \vdots \\ x_k(0) \\ \vdots \\ x_k(0) \end{bmatrix} + \begin{bmatrix} \tau_1 & \tau_1^2 & \cdots & \tau_1^z & \cdots & \tau_1^{n-1} \\ \tau_2 & \tau_2^2 & \cdots & \tau_2^z & \cdots & \tau_2^{n-1} \\ \vdots & \vdots & \ddots & \vdots & & \vdots \\ \tau_z & \tau_z^2 & \cdots & \tau_z^z & \cdots & \tau_z^{n-1} \\ \vdots & \vdots & & \vdots & \ddots & \vdots \\ \tau_{n-1} & \tau_{n-1}^2 & \cdots & \tau_{n-1}^z & \cdots & \tau_{n-1}^{n-1} \end{bmatrix} \begin{bmatrix} a_{k,1} \\ a_{k,2} \\ \vdots \\ a_{k,z} \\ \vdots \\ a_{k,(n-1)} \end{bmatrix}, \tag{28}$$

$$X_k = X_{k0} + CA_k, \tag{29}$$

where $X_k = [x_k(\tau_1), x_k(\tau_2), \cdots, x_k(\tau_{n-1})]^\top$, $X_{k0} = [x_k(0), x_k(0), \cdots]^\top$, $\top$ is the transpose operator, $A_k$ is a vector of polynomial coefficients, and $C$ is a Vandermonde matrix for the collocation nodes $\tau_1, \tau_2, \ldots, \tau_{N-1}$ without the zeroth-order terms. Its state derivative can be expressed as

$$\frac{d}{d\tau}\begin{bmatrix} x_k(\tau_1) \\ x_k(\tau_2) \\ \vdots \\ x_k(\tau_z) \\ \vdots \\ x_k(\tau_{n-1}) \end{bmatrix} = \begin{bmatrix} 1 & 2\tau_1 & \cdots & z\tau_1^{z-1} & \cdots & (n-1)\tau_1^{n-2} \\ 1 & 2\tau_1 & \cdots & z\tau_2^{z-1} & \cdots & (n-1)\tau_2^{n-2} \\ \vdots & \vdots & \ddots & \vdots & & \vdots \\ 1 & 2\tau_z & \cdots & z\tau_z^{z-1} & \cdots & (n-1)\tau_z^{n-2} \\ \vdots & \vdots & & \vdots & \ddots & \vdots \\ 1 & 2\tau_{n-1} & \cdots & z\tau_{n-1}^{z-1} & \cdots & (n-1)\tau_{n-1}^{n-2} \end{bmatrix} \begin{bmatrix} a_{k,1} \\ a_{k,2} \\ \vdots \\ a_{k,z} \\ \vdots \\ a_{k,(n-1)} \end{bmatrix}, \tag{30}$$

$$\frac{dX_k}{d\tau} = DA_k = h\frac{dX_k}{dt} = h \cdot f(X_k, U_k). \tag{31}$$

From Eqs. 29 and 31, the constraint in differential equation Eq. 24b is transcribed into the following algebraic equation:

$$X_k - x_k(0) = hCD^{-1}f(X_k, U_k). \tag{32}$$

Finally, the optimisation problem can be rewritten as follows:

$$\min_{U_k} J_{\text{nlp}} = \sum_{k=0}^{M-1} \left( \text{Expenses}_{\text{cell}}(X_k, U_k) - \text{Revenue}_{\text{cell}}(X_k, U_k) \right) \tag{33a}$$

s.t.:

$$\text{Eq. 32} \tag{33b}$$

$$X_{\text{lb}} \leq X_k \leq X_{\text{ub}} \tag{33c}$$

$$U_{\text{lb}} \leq U_k \leq U_{\text{ub}} \tag{33d}$$

$$\Delta U_{\text{lb}} \leq \Delta U_k \leq \Delta U_{\text{ub}} \tag{33e}$$

$$-\epsilon \leq X_{M-1} - X_0 \leq \epsilon \tag{33f}$$

$$-\alpha \leq P_k - P_{\text{target},k} \leq \alpha \tag{33g}$$

where Eqs. 33b–g are the equivalent constraints for the optimisation problem, Eqs. 24b–g. The constraints are imposed only on the collocation points while the cost function can be integrated numerically due to the Lobatto quadrature implemented. The optimal solution only needs to satisfy the conditions of optimality at the collocation points, and this can be efficiently computed with large scale interior point optimiser, IPOPT [48].

In this work, the optimised power profile over 48 hours (two modulation cycles) are considered. Hence, the time is partitioned into 48 segments ($M = 48$), each having 7 collocation nodes ($N = 7$), which consistently allows the solver to produce a feasible solution. The segmentation and collocation nodes are illustrated in Figure 6.

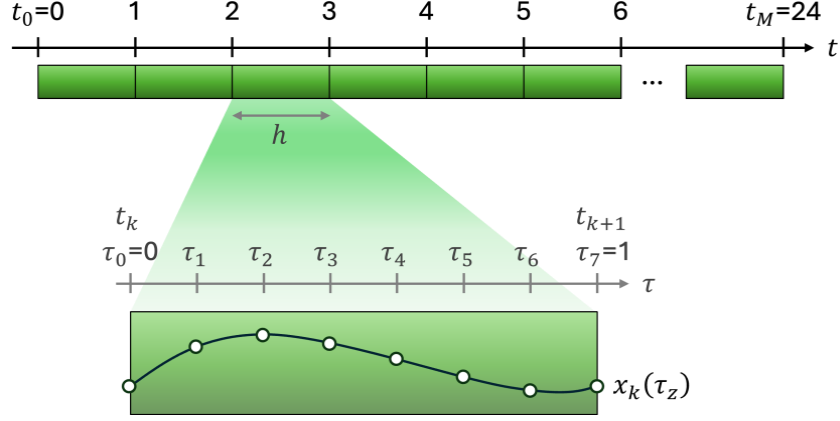

Figure 6: Optimisation window partitioned into 24 segments, each with 7 collocation nodes.

### 3.2.3 Feedback-Update Strategy

As larger reduction cells are built, the process becomes increasingly uneven in terms of material and energy distribution throughout the cell. The spatially distributed model captures the coupled material and thermal dynamics with high spatial resolution, but this also results in the number of nonlinear programming constraints growing proportionally to the product of the number of states and collocation points, leading to scalability challenges. While the reduced-order model significantly reduces computational requirements, it introduces errors due to variations in spatial system properties, such as the thermal inertia illustrated in Figure 5. To address this, we introduce a novel feedback-update optimisation strategy with shrinking time horizon to compute the optimal line current and ACD trajectories (*i.e.*, power consumption control profile). It is emphasised that the scope of this study is to perform an offline analysis of the economic feasibility and process operability over a period of two modulation cycles (48 hours).

Figure 7 illustrates the proposed optimisation framework based on shrinking optimising horizons. The nonlinear programming problem defined in Eq. 33, constructed using a reduced-order model of the cell, is initially solved over the entire 48-hour horizon. Leveraging this model that predicts cell dynamics, this yields an optimal trajectory $u^*(t)$, comprising the line current and ACD, which satisfies all operational constraints. To mitigate the impact of model-reduction error, the power consumption control profile of the first $\theta$ time ($\theta = 10$ min for the studies in Section 4),

$$u^{*(0)}(t) := u^*(t), \quad t \in [0, \ \theta], \tag{34}$$

is applied to the spatially distributed model in an open-loop simulation, providing a more accurate prediction of the system's state trajectory. The terminal state from the spatially distributed model $x(\theta)$ is then spatially averaged using mass-based weighting, ensuring consistency with the reduced-order model representation. This terminal state is used to initialise the optimiser at time $\theta$ and the optimisation is re-solved over the remaining time horizon, $t \in [\theta \quad t_f]$. This process is repeated, with the optimal trajectory

$$u^{*(j)}(t) := u^*(t), \quad t \in [j\theta, \ (j+1)\theta] \tag{35}$$

retained and implemented on the spatially distributed model with each $j$-th iteration. The optimisation horizon shrinks at each iteration until the full 48-hour period has been traversed.

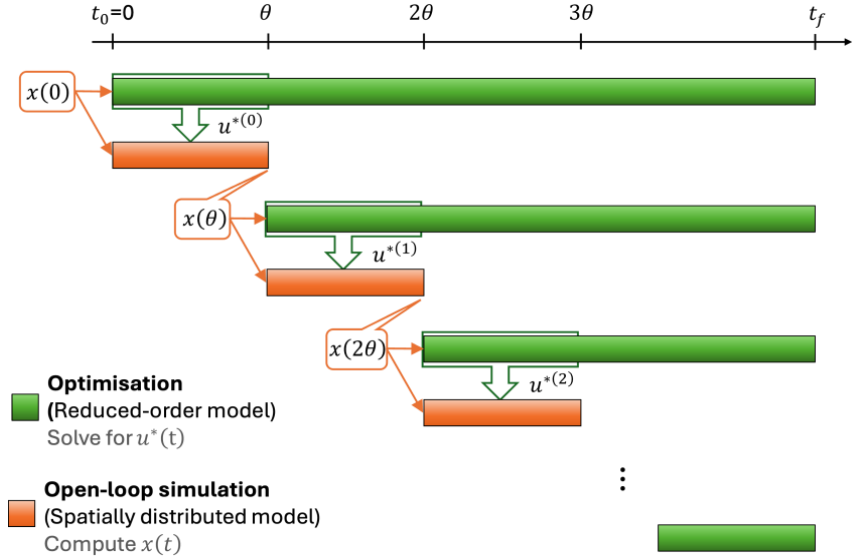

Figure 7: Feedback-update optimisation approach with shrinking optimisation horizons and fixed simulation horizons $\theta$.

This approach retains the planning benefits of long-horizon optimisation while utilises a spatially distributed model to provide state feedback to increase robustness of the offline model prediction. By periodically updating the process variables in the reduced-order model with more accurate state information obtained from the spatially distributed simulation, the optimisation solution more accurately reflects the spatially distributed, coupled mass and thermal dynamics of the full-order system.

Optimisation in tandem with the feedback-update framework forms the basis of an application which can yield feasible line current and ACD profiles based on technology specific models to improve smelter profitability whilst ensuring the process remains within strict operating constraints. This tool enables the specification of set power profiles as optimisation constraints or may alternatively take inputs of power price profiles to determine near optimal operational response, yielding dynamically optimised strategies to extract maximum economic benefit whilst adhering to the strict operational constraints of aluminium reduction technology.

## 4. Optimisation Studies

The models described in Section 3.1 were simulated with the feedback-update optimisation described in Section 3.2 to study the potential benefits of three different power modulation scenarios, while ensuring the cells are working within acceptable operating region. In these scenarios, the cells are subjected to the varying power input, and the cell dynamics are evaluated to yield insight into potential operational issues and to assess the feasibility of the power modulation regimes. In each scenario, we consider an optimisation time horizon of 48 hours to capture the long-term cell behaviour during power modulation whilst also limiting the computational costs arising from long horizon window.

The first scenario considers a changing power setpoint of every 12 hours in a diurnal load shifting pattern [49]. This benefits smelters by honouring contractual obligations with the grid to reduce power in times of peak load demand. This allows smelters to provide demand-side response services to the energy grid, with financial incentives being rewarded through these agreements [5, 49]. As these financial agreements are commercial in confidence, their financial gains are not assessed in this paper.

The second and third modulation scenarios consider a time-varying power pricing, and the input power is allowed to respond freely to maximise financial gains. The second scenario considers a time-of-use tariff profile typical in NSW, Australia, while the third scenario considers power spot prices published by the Australian Energy Market Operators (AEMO) [50]. Spot pricing can fluctuate drastically, representing the need for more extreme cases of power modulation. This evaluates the potential economic benefits of permitting fast smelter load response to highly variable spot prices.

**4.1 Scenario I: 20% Diurnal Load Shifting Power Variation**

Diurnal load shifting [49] exploits the daily variations of power demand in the grid by decreasing smelter power demand to save on smelter power costs and then increase aluminium production in times of low grid demand. It can be implemented as a percentage decrease in power during the day and percentage increase of the same magnitude during the night; a periodic power consumption repeated every 24 hours.

The cell technology of this study operates with a line current of 425 kA and an average ACD of 2.8 cm under nominal conditions. A ±20% power variation is introduced during the day and night, to demonstrate the need for drastic ACD changes to cope with the more prominent heat fluctuations and the associated threat to cell thermal balance. The power profiles were first-order filtered using a time constant of 0.2 h to ensure the system dynamics are well-behaved with finite derivatives at the power transitions. The optimiser solved the problem, and Figure 8 shows the power profile achieved, along with the required ±20% variation. Figure 9 shows the optimised line current and ACD trajectories that achieve this target. One of the key process states, ledge thickness, is also included in the figure to illustrate the thermal capacity of the cell.

During the set period of increased power, the optimiser ramped the line current up and the ACD down to the lower bound. From the cost function formulation, the optimisation algorithm seeks to maximise cell profits by maximising aluminium metal production, and this can be achieved with higher line current as explained via Faraday's law of electrolysis (Eq. 3). At the same time, ACD was decreased to lower electrolyte voltage drop, limiting the heat generation to adhere to thermal balance constraints. This also permits greater line currents to maximise profits for a given power profile.

Increasing line current while squeezing ACD is already a common industrial practice when determining a new steady-state operation at higher power input levels. However, our optimisation approach considers and anticipates the process dynamics, and thus can produce time-series line current and ACD profiles that meet the power variation objective while maximising profits and remaining within safe operating regimes. For example, apart from ACD, line current variations also lead to changes in cell voltage due to linear ohmic and nonlinear overpotential components of the cell voltage. Yet, the optimiser can determine profiles for both ACD and line current variations that are compatible. We note in the final 3 hours of the simulation, an increased ACD also ensures the ledge returns to the same value as its initial condition by remelting the ledge, hence the effects of power modulation are reversible with arbitrary periodicity of 48-hours.

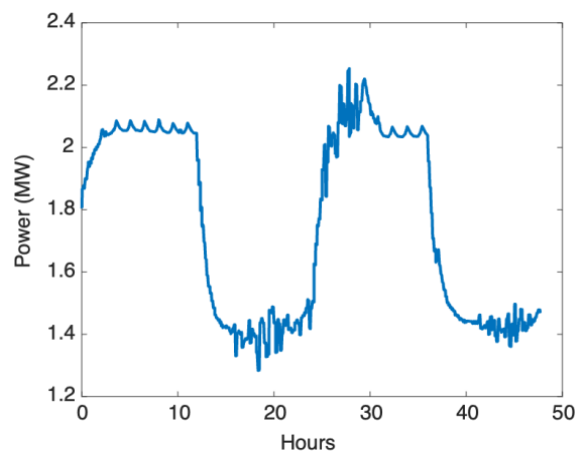

Figure 8: 20% diurnal load shifting profile for the aluminium reduction cell.

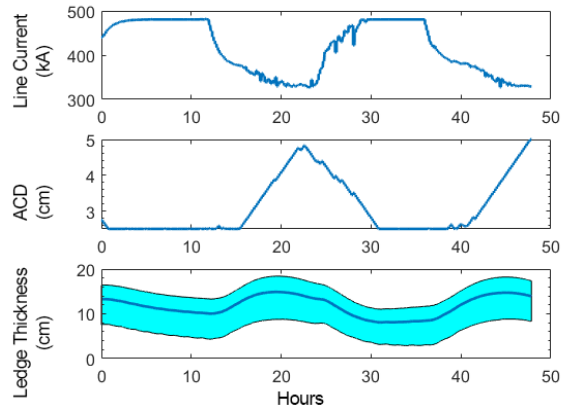

Figure 9: Optimal Line Current and ACD profiles with ledge thickness evolution over time, 20% diurnal load shifting.

It is apparent that the minimal ledge thickness is close to the allowed lower bound of 2 cm, specifically between the hours of 30 to 40. Due to a reduced flexibility in operation during periods of increased power (due to reductions in ACD below the minimum contributing to voltage noise and pot instability), given the set power consumption it becomes apparent that a ±20% diurnal load shifting approaches the physical limit for the modelled cell technology. This is further supported by the consistent infeasibility of solution in the optimisation problems when the load shifting percentages were raised to ±25% and higher for this cell technology.

Despite the requirement for ACD changes and the consideration of changing species concentrations with varying electrolyte mass due to ledge formation/melting, the simulation portrays reasonable cell voltage trends, confined between approximately 3.8 V to 4.8 V as seen in Figure 10.

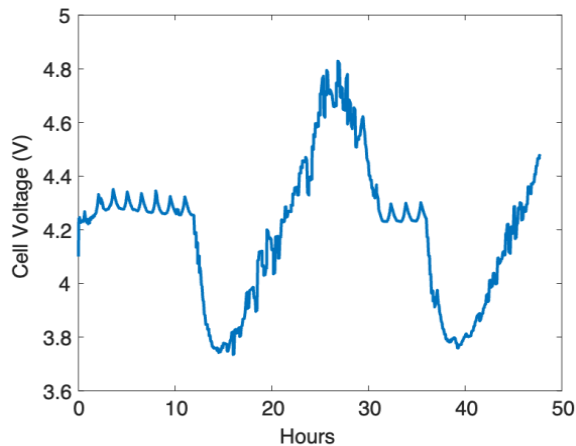

Figure 10: Cell Voltage, 20% diurnal load shifting.

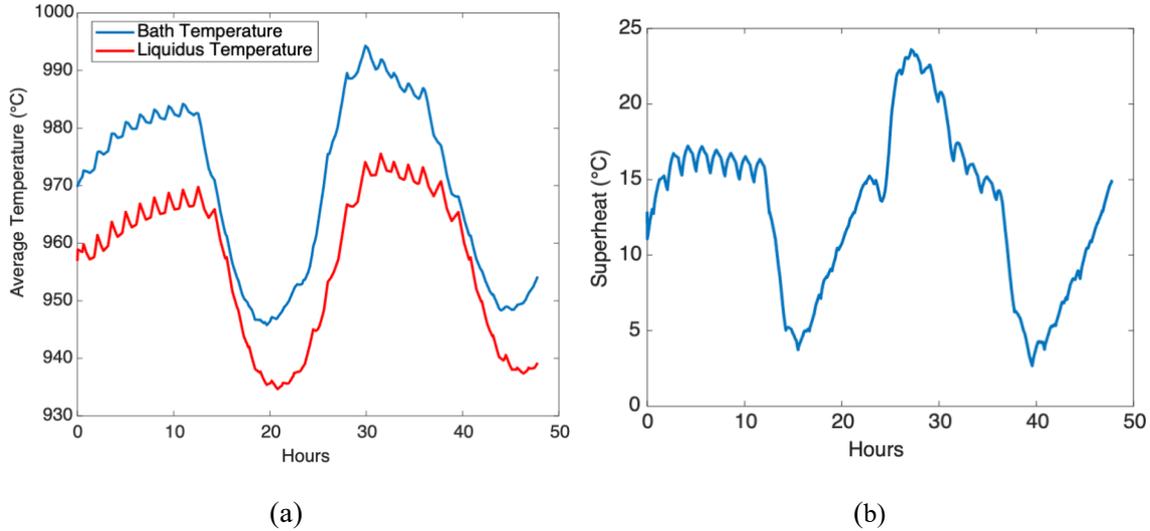

(a)                          (b)

Figure 11: Variation in (a) average electrolyte/bath temperature (°C) and liquidus temperature, (b) cell superheat with time, 20% diurnal load shifting.

Figure 11(a) shows that the electrolyte temperature varies between approximately 945 °C and 995 °C. The liquidus temperature, which defines the temperature at which the cell electrolyte becomes completely liquid, also varies due to the changes in electrolyte composition brought about by the ledge formation/melting. Meanwhile, Figure 11(b) presents the changes in superheat (the difference between electrolyte and liquidus temperatures) with a maximum of approximately 24 °C and a minimum of approximately 4 °C. The magnitude of superheat represents two key aspects in aluminium smelting: firstly, it is a representation of the availability of energy to dissolve feed materials, such as alumina fed to the cell, and secondly, the superheat provides an indication of the heat transfer through the sides of the reduction cell, determining the degree of cryolitic ledge melting/freezing [51]. Figure 11(b) demonstrates dynamic changes in liquidus temperature lagging the changes in electrolyte temperature, in turn causing fluctuations in superheat throughout the load shifting period. This can be explained by liquidus temperature being solely dependent upon the chemical composition of the electrolyte, with the relatively slower variations in electrolyte total mass (as evidenced by slow ledge dynamics in Figure 9) creating this lag effect. Whilst the regime presented is feasible with adherence to constraints on ledge thickness, larger superheats are reported in literature to contribute towards a more significant reduction in current efficiency, and hence increase energy requirements whilst reducing aluminium production [52]. As such, an ideal control philosophy in a power modulation context would require consideration of both thermal properties and electrolyte composition together to effectively regulate superheat. The extent of these current efficiency reductions is based on empirical relations specific to cell technologies, and since the relation has yet to be obtained for the specific technology of this study, variations in current efficiency have been excluded from the formulation of the optimisation problem.

Of concern is the high magnitude of superheat during the second increased power period between 24-th to 36-th hour. As previously mentioned, increased superheats correspond to decreased current efficiency hence decreased aluminium production [52]. Under the current formulation of the optimal control problem, there is a clear exchange between current efficiency and ensuring that thermal balance constraints of the aluminium reduction cell are met. Hence, these results emphasise the importance of considering variable current efficiencies in the aluminium reduction cell model, motivating further studies to attain the empirical model for current efficiency for the specific cell technology, incorporate this relation into the dynamic model and reformulate the optimal control problem under this consideration.

## 4.2 Scenario II: Time-of-use Tariff Scenario

Optimal regimes in response to more general, time-of-use tariff scenarios were studied with the feedback-updated method using the spatially distributed model (see Figure 12). A typical time-of-use tariff profile as used by NSW energy providers in Australia was input to the optimisation problem to assess feasibility and profitability of operating a smelter cell with variable electricity pricing contracts. Note in the following simulation, power is permitted to float freely given determined changes in ACD and line current.

Optimisations were conducted using two sets of constraints for the line current rate of change, to assess the impact of fast current changes (±360 kA/h) and slower changes (±36 kA/h). Under the ±360 kA/h constraint, a comparison of Figure 12, Figure 13, and Figure 14 show how the optimiser takes advantage of periods with low power prices to generate sufficient heat, ensuring that the average ledge thickness remains below the specified upper bound of 15 cm. During these periods of low power cost, the line current can be rapidly increased while the anode-cathode distance (ACD) is relatively large (around 4.5 cm), allowing significant heat generation within the cell electrolyte. The profiles are optimised for economic cost, with the line current being used to quickly modulate power and heat generation, while the ACD changes more gradually, as illustrated in Figure 14.

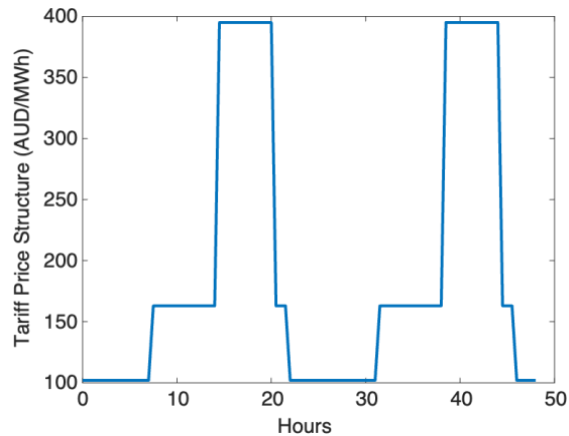

Figure 12: Time-of-use tariff over 48 hours.

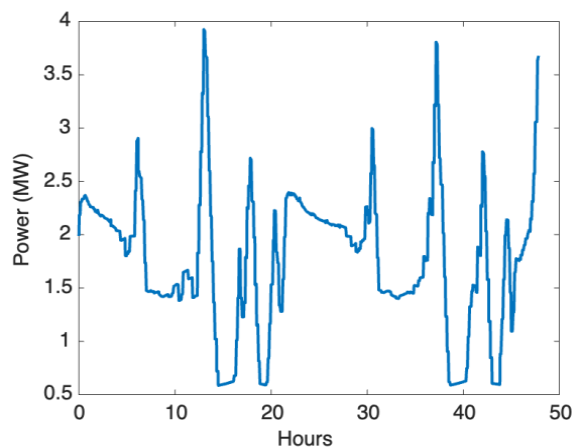

Figure 13: Optimised power demand profile in response to time-of-use tariff between ±0.36 cm/h for ACD and between ±360 kA/h for line current.

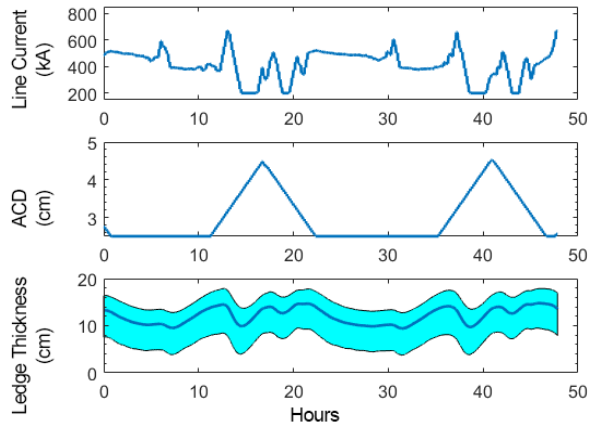

Figure 14: Optimised line current and ACD profiles with evolution of ledge thickness with time in response to time-of-use tariff between ±0.36 cm/h for ACD and between ±360 kA/h for line current.

Figure 15 and Figure 16 show the same scenario under stricter rate of change constraints for the line current (±36 kA/h) still presents an economically optimal regime, however economic benefit suffers when compared with the previous case. In this scenario, the ledge is confined within a tighter range associated with the limitation of line current rate of change. Whilst the constraints are held, there is reduced opportunity for economic benefit due to slower transitions in power consumption, challenging the ability to respond to changes in the power prices. This demonstrates the impact of physical limitations on the potential smelter economic benefit, *i.e.*, permitting sharper changes in power consumption allows for longer operation near the constraint boundaries where economic benefit is most significant.

Figure 17(a) compares the single cell profit across both sets of constraints for the line current rate of change (±360 kA/h and ±36 kA/h) against nominal operation. This was calculated with the cost function defined in Eq 24, which accounts for expenses related to costs of raw materials and electricity, and revenue from the market value of aluminium metal produced, while excluding capital, maintenance, and labour costs. Figure 17(b) shows a consistent improvement in profitability with power modulation, with higher permissible rate-of-change in current being more effective. Power modulation enables rate of aluminium production to be adjusted flexibly according to power costs.

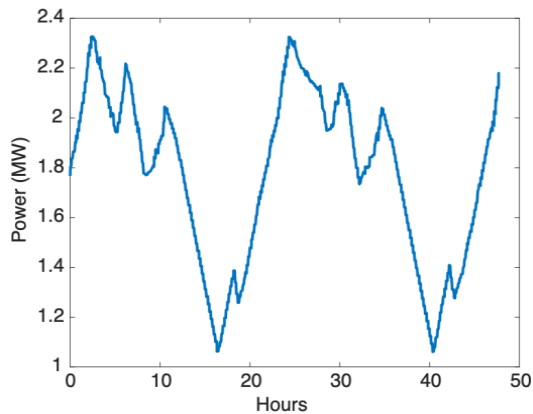

Figure 15: Optimised power demand profile in response to time-of-use tariff between ±0.36 cm/h for ACD and between ±36 kA/h for line current.

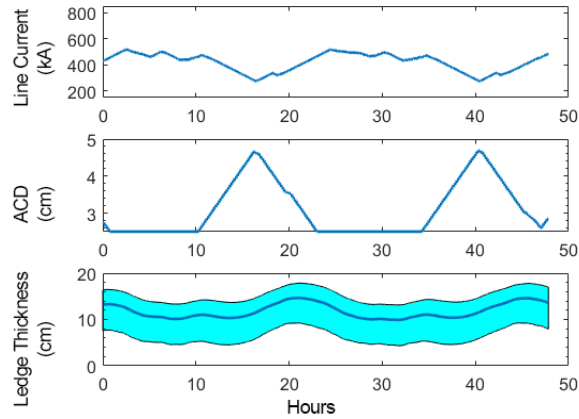

Figure 16: Optimised line current and ACD profiles with evolution of ledge thickness with time in response to time-of-use tariff between ±0.36 cm/h for ACD and between ±36 kA/h for line current.

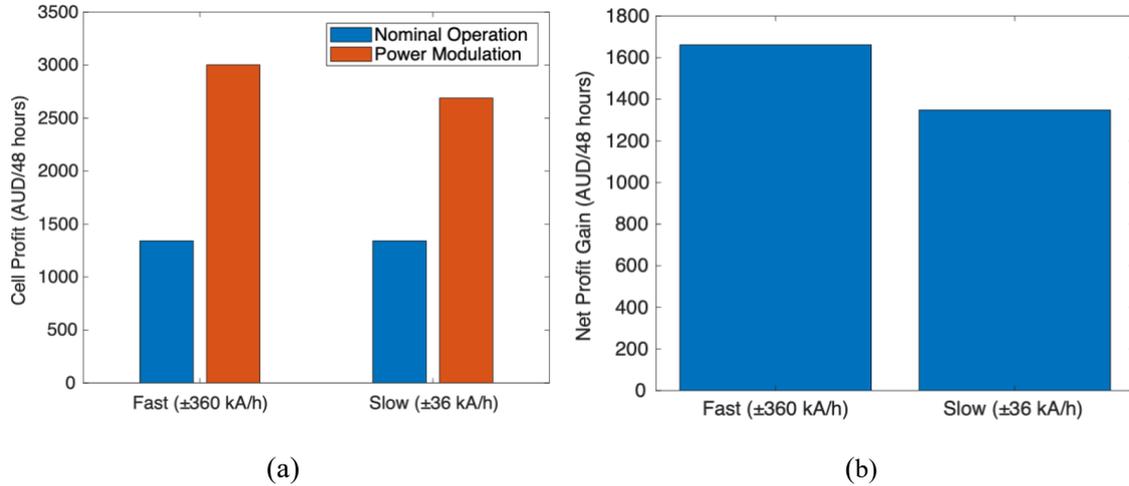

(a) (b)

Figure 17: Comparison of (a) cell profits for nominal operation and power modulation (normal and slow response cases) in a 48-hour period and (b) net profit gained for power modulation vs nominal operation.

### 4.3 Scenario III: Power Spot Price Profile

As an extension to Scenario II (time-of-use tariff studies), optimisations were performed under consideration of NSW Australia spot market prices, permitting the cell power load to respond freely to maximise cell profits in a 48-hour period. Typically, the aluminium smelting industry relies upon agreements with the grid such that power prices are consistent and manufacture remains economically viable for constant smelter power loads. Whilst these studies represent an extreme power modulation scenario, they provide insight into efficacy of power modulation for economic benefit and reduction of costs associated with spot price pricing scheme. Evaluation of such pricing schemes yield valuable insights, particularly when considering the uncertain future of power pricing for aluminium smelters as the present agreements come to an end.

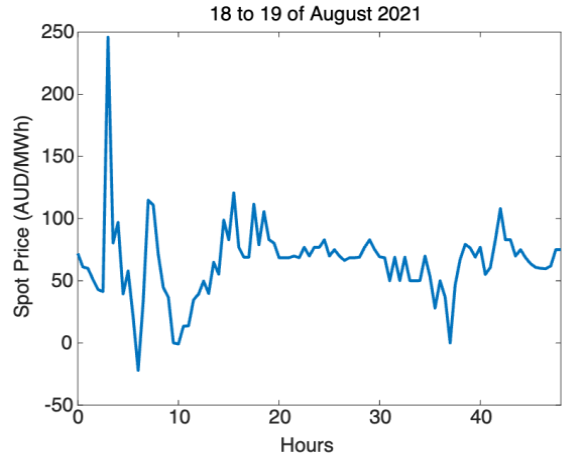

Figure 18: NSW Australia power spot price profile starting at midnight for 18th to 19th of August 2021.

Optimisation was performed considering a 48-hour window subject to the previous constraints detailed for ledge thickness (minimum ledge thickness greater than or equal to 2 cm, average ledge thickness less than or equal to 15 cm) and constraints on the line current and ACD rates of change (bounds of ±360 kA/h and ±0.36 cm/h respectively). Using the feedback-update optimisation, power modulation was optimised on a monthly frequency of 48-hour spot price profiles from August 2021 to January 2022, producing the overall cell profits for these periods and compared with the profits under nominal operation of 425 kA line current and an average ACD of 2.8 cm. A sample spot price profile for August 2021 is given in Figure 18, showing the typical spot price fluctuations observed in NSW.

Figure 19(a) portrays the calculated single cell profit over 48-hour periods, chosen for large fluctuations observed in the spot price, in different months for both nominal operation and power modulation cases. Figure 19(b) demonstrates consistent improvements across all months with power modulation, averaging an improvement of A$1,232. As noted previously, this does not include costs related to capital, maintenance, and labour.

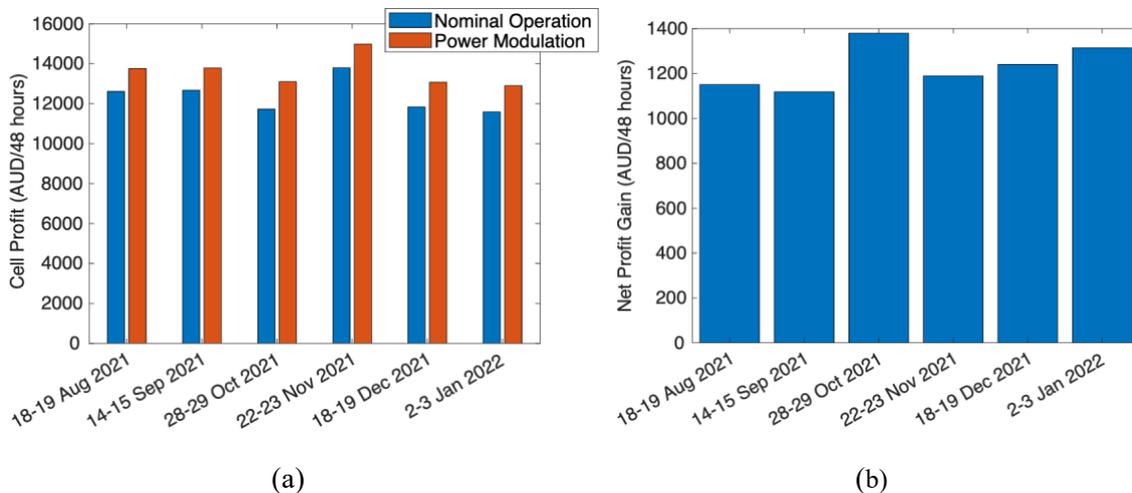

(a) (b)

Figure 19: Comparison of (a) single cell profits during nominal operation and power modulation under power spot prices for various months and (b) net profit gained for power modulation vs nominal operation.

## 5. Conclusions and Future Work

In this paper, the operability and economic benefits of Hall-Héroult cells under power modulation was studied as an optimisation problem with an economic cost function. The multi-timescale nature of the process was addressed by using continuous-time model with orthogonal collocation methods applied to efficiently solve the optimisation problem with nonlinear programming solver. Errors arising from the use of reduced-order model to predict process dynamics of a spatially distributed nature is addressed by incorporating a high spatial fidelity model in a feedback loop. Overall, this strategy allows the power profiles (line current and ACD trajectories) to be determined which maximises smelter profits while ensuring the cell remains within strict operating constraints of the process.

The studies identified an approach towards the maximum feasible ±20% diurnal load shifting for the cell technology studied, primarily due to the ACD lower bound of 2.5 cm paired with extended periods of minimum ledge thickness at the lower boundary of 2 cm. Since ACD below this limit is considered unfavourable due to potential issues with cell voltage noise and pot instability, this implies strict power constraints will contribute a minimum ohmic heat generation in the cell. This may in turn render operation infeasible at higher power inputs due to complete melting of the ledge at points in the cell, damaging the refractory cell wall and shortening cell lifespan.

Optimal regimes in response to a time-of-use tariff were also studied to assess the operational flexibility of aluminium smelter cells, where power consumption was permitted to float freely. The results demonstrate how ACD can be raised slowly and pre-emptively in preparation for a substantial decrease in power, hence ensuring the conservation of thermal balance. Further, comparisons between power modulation and nominal operation for spot market power prices demonstrate the potential economic benefits of power modulation under spot price power pricing structure, with an average increase in profits of A$1,232 per 48 hours per cell for the studied cell technology. This figure may vary in practice when implemented as an online control, since power price may vary in real-time. Participation in grid power balancing services also offers an additional stream of financial incentives.

To further improve the optimiser results, future works include accounting for variable current efficiency and incorporating renewable energy storage into the considerations, permitting insights into feasible operating regimes of aluminium reductions cells in conjunction with appropriate sizing of battery farms to improve process economic viability and the reduction smelter carbon footprints. The use of heat exchangers with variable heat transfer rate can also be considered in conjunction with dynamically changing ACD and line current to achieve a larger modulation range. Additionally, whilst studies were performed for a single cell, in industry multiple cells are connected in series, with each cell sharing the same line current with any changes in line current affecting all the connected cells. This warrants further studies to determine optimal line current and ACD trajectories with respect to a multiple cell distributed system.

This work forms the preliminary structure for further studies into online control policies of aluminium reduction cells, with an intention to extend the optimisation for application in an economic-model-predictive-control (EMPC) framework. Prior to the online control application, further studies must be conducted into potential measurements attainable from smelter cell technology using a combination of manual and automatic measurements, as well as the application of state observers such that full-state feedback may be utilised. In essence, observation of the cell states, particularly the ledge thickness, presents as one of the next major challenges in improving the autonomy of the Hall-Héroult process.

## 6. Acknowledgements

The authors acknowledge the financial support from Australian Research Council (ARC) Research Hub for Integrated Energy Storage Solutions (IH180100020), and Emirates Global Aluminium (EGA). The technical support from EGA Jebel Ali Operations, especially the Technology Development and Transfer Team and Operations Team, is greatly appreciated. The authors also wish to thank Michel Reverdy and Vinko Potocnik for their useful comments.